\documentclass[nofootinbib,aps,prb,reprint,superscriptaddress,twocolumn, draft=false]{revtex4-1}
\pdfoutput=1 
\usepackage{hyperref}
\hypersetup{colorlinks=true,
		    linkcolor=blue,
		    citecolor=blue,
		    allcolors=blue}
\usepackage{amsmath, amsfonts, bm, bbm, braket}
\numberwithin{equation}{section}
\renewcommand\theequation{\arabic{section}.\arabic{equation}}
\usepackage{graphicx}
\usepackage{float}
\usepackage{natbib}
\usepackage{lipsum}	
\usepackage{comment}	    
\newcommand{\norm}[1]{\left\lVert#1\right\rVert}
\renewcommand{\d}{\mathrm{d}}
\renewcommand{\Im}[1]{\mathrm{Im}(#1)}
\renewcommand{\Re}[1]{\mathrm{Re}(#1)}
\newcommand{\id}{\mathbbm{1}}
\renewcommand{\H}{\mathcal{H}}
\newcommand{\E}{\epsilon}
\newcommand{\w}{\omega}
\newcommand{\A}{\mathcal{A}}

\newcommand{\G}{\mathcal{G}}
\newcommand{\M}{\mathcal{M}}
\renewcommand{\k}{\bm{k}}
\newcommand{\q}{\bm{q}}

\newcommand{\vp}{\varphi}

\begin{document}

\title{Selective Phonon Damping in Topological Semimetals}

\author{Jacob S. Gordon}
\affiliation{Department of Physics, University of Toronto, Ontario M5S 1A7, Canada}
\author{Hae-Young Kee}
\email{hykee@physics.utoronto.ca}
\affiliation{Department of Physics, University of Toronto, Ontario M5S 1A7, Canada}
\affiliation{Canadian Institute for Advanced Research, CIFAR Program in Quantum Materials, Toronto, ON M5G 1M1, Canada}

\date{\today}

\begin{abstract}
Topological semimetals are characterized by their intriguing Fermi surfaces (FSs) such as Weyl and Dirac points, or nodal FS, and their associated surface states. Among them, topological crystalline semimetals, in the presence of strong spin-orbit coupling, possess a nodal FS protected by non-symmorphic lattice symmetries. In particular, it was theoretically proposed that $\mathrm{SrIrO}_{3}$ exhibits a bulk nodal ring due to glide symmetries, as well as flat two-dimensional surface states related to chiral and mirror symmetries. However, due to the semimetallic nature of the bulk, direct observation of these surface states is difficult. Here we study the effect of flat-surface states on phonon modes for $\mathrm{SrIrO}_{3}$ side surfaces. We show that mirror odd optical surface phonon modes are damped at the zone center, as a result of coupling to the surface states with different mirror parities, while even modes are unaffected. This observation could be used to infer their existence, and experimental techniques for such measurements are also discussed.
\end{abstract}

\maketitle

\section{\label{sec:introduction}Introduction}
Topological semimetals (TSMs) are materials with non-trivial crossing of valence and conduction bands at points, lines, or loops within the Brillouin zone. They are topologically protected in the sense that the crossing cannot be lifted by a symmetry-preserving perturbation. The symmetry which protects these crossings can be global, such as time-reversal, or crystalline, and are classified through a bulk topological invariant of the Bloch states in a neighbourhood of the crossing. Proposals for TSMs so far include Weyl points\cite{murakami2007phase, murakami2008universal, wan2011topological, burkov2011weyl, xu2011chern, fang2012multi, yan2017topologicalwsm} in systems with broken inversion or time-reversal symmetry, Dirac points\cite{young2012dirac, wang2012dirac, armitage2018weyl}, as well as nodal line semimetals\cite{burkov2011topological, carter2012semimetal,fang2016topologicalnlsm}. In addition, the non-trivial bulk topology in TSMs may manifest itself through associated surface states, including Fermi arcs between bulk Weyl or Dirac points.

Among these TSMs, topological crystalline semimetals (TCSMs) with strong spin-orbit coupling possess a nodal FS which is protected by non-symmorphic lattice symmetries. In particular, it has been proposed theoretically that SrIrO$_3$ exhibits a nodal ring FS\cite{carter2012semimetal} protected by two perpendicular glide symmetries\cite{fang2015tnlsm, chen2016tcsm}. This nodal line FS is interesting, as it may act as a parent state for other nodal FS structures when symmetry-breaking perturbations are added. Associated with the bulk topology, protected by glide symmetries, are double helicoid surface states\cite{fang2016topological, chen2016tcsm} on the $(001)$ top surface. Furthermore, on side surfaces perpendicular to $(001)$ flat two-dimensional surface states associated with mirror and chiral symmetries are predicted\cite{chen2015topological, kim2015surface} to exist. 

Apart from difficulty in the synthesis of SrIrO$_3$\cite{longo1971structure}, these flat 2D surface states on side surfaces are difficult to observe directly due to the fact that the bulk is semimetallic. Resistivity measurements\cite{matsuno2015engineering} performed on thin films, synthesized using pulsed laser deposition, as well as ARPES\cite{nie2015interplay} measurements confirm the semimetallic nature and the nodal FS. However, a clear signature of the flat side surface states remains elusive. 

In this work, we propose that phonon modes can be used to infer the existence of side surface states. Symmetry properties of the surface state wave function leads to a unique electron-phonon coupling which will damp only certain optical phonon modes at the zone center.  This paper is organized as follows. In Section~\ref{sec:surface-states} we study the side surface states in detail through an analytic solution of the wave function (\ref{subsec:open-boundary-wavefunction}), and direct numerical diagonalization (\ref{subsec:slab-calculations}). In Section~\ref{sec:epi-polarization} we show how the symmetry of the electronic wave function constrains the electron-phonon interaction, and calculate the first order phonon self-energy (\ref{subsec:density-response}, Appendix~\ref{app:bubble-calculation}). Finally, experimental techniques used to measure this effect are discussed in Section~\ref{sec:discussion}. 

\section{\label{sec:surface-states}Surface States in a TCSM}

\begin{figure}
	\includegraphics[width=0.75\linewidth]{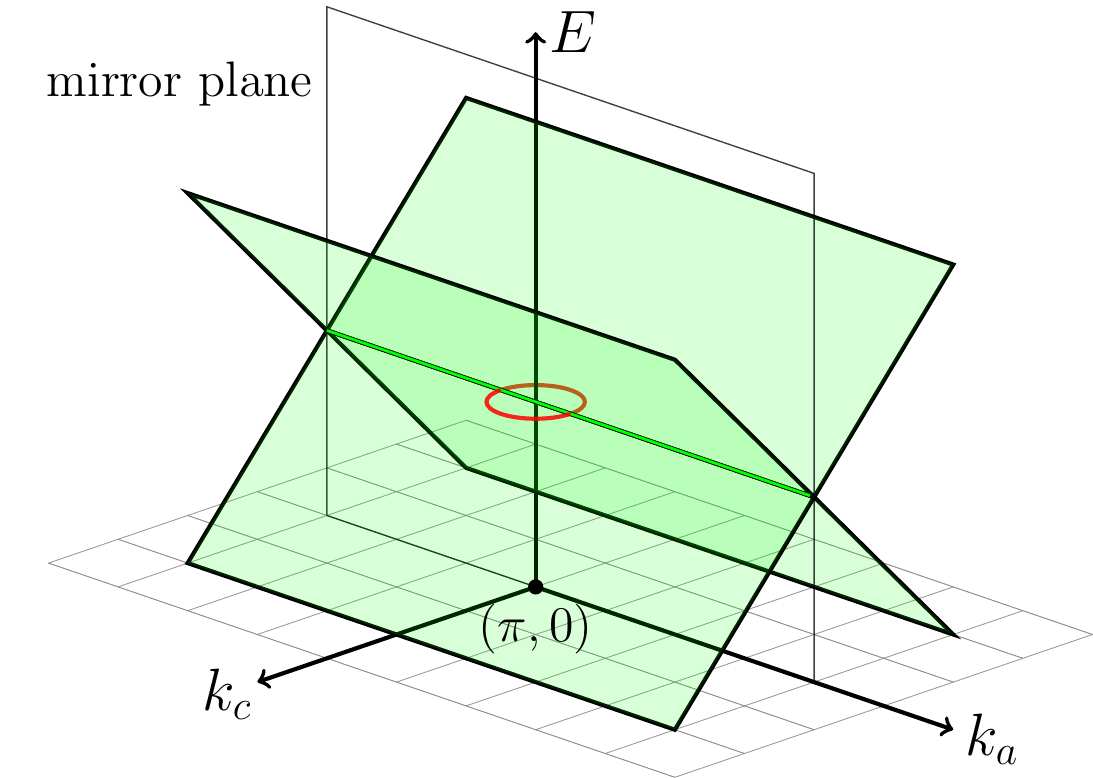}
	\caption{(colour online)  Surface states in AIrO$_3$ for a $\hat{\bm{b}}$ crystal termination. The two branches, shown in green, disperse linearly away from $k_c = \pi$ but are flat in the $k_a$ direction, forming a line of 1D Dirac cones. Projection of the bulk nodal ring at $(k_c,k_a) = (\pi,0)$ is shown as a red ellipse.}
	\label{fig:SBZ}
\end{figure}

Through a combination of strong spin-orbit coupling and crystal field splitting, $j_{\mathrm{eff}} = \tfrac{1}{2}$ states provide a good basis for a low-energy description\cite{carter2012semimetal,kim2015surface} of orthorhombic perovskite iridates AIrO$_3$ (A an alkaline earth metal, space group \textit{Pbnm}). The unit cell of AIrO$_3$ contains four Ir atoms ($B,R,Y,G$) on which the $j_{\mathrm{eff}}$ states live, and are distinguished by distortion of the surrounding oxygen octahedra. The full tight-binding Hamiltonian, derived in Ref. [\onlinecite{carter2012semimetal}], is written in the basis
\begin{equation}\label{eq:tb-basis}
	\psi = (c_{B\uparrow},c_{R\uparrow},c_{Y\uparrow},c_{G\uparrow},c_{B\downarrow},c_{R\downarrow},c_{Y\downarrow},c_{G\downarrow})^T,
\end{equation}
where $\uparrow,\downarrow$ refer to $j_{\mathrm{eff}}^z = \pm \tfrac{1}{2}$. It takes the form
\begin{align}\label{eq:full-tb}
\begin{split}
\H_{\k} &= \hspace{2.5mm} \Re{\E^p_{\k}}\tau_x + \Im{\E^p_{\k}}\sigma_z\tau_y + \E^z_{\k}\nu_x \\
&\hspace{3mm}+\Re{\E^d_{\k}}\nu_x\tau_x + \Im{\E^d_{\k}}\nu_y\tau_y\\
&\hspace{3mm}+[\Re{\E^{po}_{\k}}\sigma_y + \Im{\E^{po}_{\k}}\sigma_x]\nu_z\tau_y \\
&\hspace{3mm}+[\Re{\E^{zo}_{\k}}\sigma_y + \Im{\E^{zo}_{\k}}\sigma_x]\nu_y\tau_z \\
&\hspace{3mm}+[\Re{\E^{do}_{\k}}\sigma_y + \Im{\E^{do}_{\k}}\sigma_x]\nu_x\tau_y \\
&\hspace{3mm}+[\Re{\E^{d1}_{\k}}\sigma_y + \Im{\E^{d1}_{\k}}\sigma_x]\nu_y\tau_x,
\end{split}
\end{align}
where three sets of Pauli matrices correspond to pseudospin ($\bm{\sigma}$), layer ($\bm{\nu}$), and in-plane sublattice ($\bm{\tau}$). Tight-binding parameters and the form of the functions $\E_{\k}$ are listed in Appendix~\ref{app:tb-hamiltonian}. Next-nearest neighbour hopping ($\lambda_{\k}$) has been left out for simplicity, but its effect is discussed in Section \ref{sec:epi-polarization}. This model exhibits an elliptical one-dimensional (1D) FS called the nodal ring, consistent with \textit{ab initio} calculations\cite{carter2012semimetal} and protected by non-symmorphic symmetries\cite{fang2015tnlsm, chen2016tcsm}.

This topological crystalline semimetal also exhibits\cite{chen2015topological} a pair of surface zero modes on the mirror-symmetric line $k_c = \pi$ for all side surfaces except those perpendicular to a weak index $\bm{M} = \hat{\bm{a}} + \hat{\bm{b}} \parallel \hat{\bm{y}}$. These modes are protected by a combination of mirror symmetry 
\begin{equation}\label{eq:mirror-symmetry}
\Pi_m = i\sigma_z\nu_x \hspace{5mm}(k_a,k_b,k_c) \mapsto (k_a,k_b,-k_c),
\end{equation} 
and an emergent chiral symmetry 
\begin{equation}\label{eq:chiral}
	\mathcal{C} = \sigma_z\nu_y\tau_z,
\end{equation}
which anti-commutes with the Hamiltonian Eq.~\ref{eq:full-tb} on the $k_c = \pi$ plane. Zero modes are indicated in Fig.~\ref{fig:SBZ} for a $\hat{\bm{b}}$ crystal termination with a thick line. Away from $k_c = \pi$ the surface modes disperse linearly, forming a line of 1D Dirac cones which are flat in the $k_a$ direction. Time-reversal symmetry is of the usual form
\begin{equation}
	\mathcal{T} = i\sigma_y\mathcal{K} \qquad  \k \mapsto -\k,
\end{equation}
where $\mathcal{K}$ denotes complex conjugation. 

While the dispersion of the surface states is well understood, little is known about their wave functions which must be known to describe their response to density perturbations. Therefore, in this section we describe their wave functions through two complementary approaches. First, we obtain the wave function analytically at $k_c = \pi$ by solving an open boundary problem with the bulk Hamiltonian. Second, we extend this solution away from $k_c = \pi$ by numerically diagonalizing the Hamiltonian in a slab geometry. Finally, we show how one can form total and `relative' density combinations which will couple to certain phonon modes discussed in Section~\ref{sec:epi-polarization}. 
 
\subsection{\label{subsec:open-boundary-wavefunction} Open Boundary Wave Function at \texorpdfstring{$\bm{k_c = \pi}$}{$k_c = \pi$}}

\begin{figure}
	\includegraphics[width=0.8\linewidth]{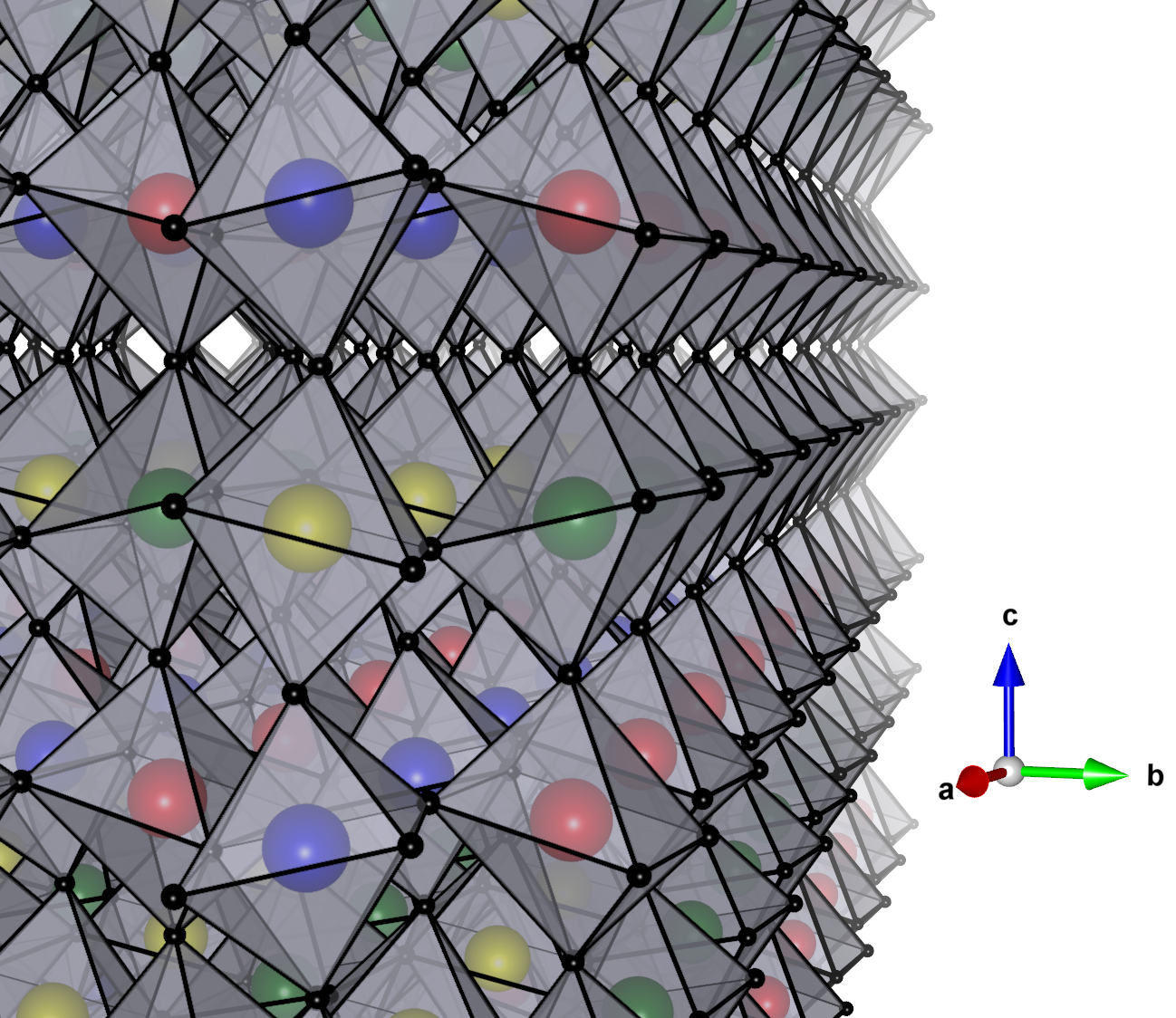}
	\caption{(colour online) Slab geometry of AIrO$_3$ with a $\hat{\bm{b}}$ crystal termination which is periodic in the $\hat{\bm{a}}$ and $\hat{\bm{c}}$ directions. Since $x_b < 0$ is the bulk, the red and green iridiums are closest to the edge at $x_b = 0$. }
	\label{fig:slab-geometry}
\end{figure}

For definiteness, we focus on the open boundary problem for a $\hat{\bm{b}}$ surface in the half space $x_b < 0$, with $x_b > 0$ the vacuum as shown in Fig.~\ref{fig:slab-geometry}. Periodic boundary conditions are imposed in the $\hat{\bm{a}},\hat{\bm{c}}$ directions, and with this crystal termination the red and green iridiums lie on the surface. Starting with the Hamiltonian Eq.~\ref{eq:full-tb}, we fix $k_c = \pi$ so that both mirror and chiral symmetries are present. On this mirror-symmetric plane, the mirror operator Eq.~\ref{eq:mirror-symmetry} simplifies to
\begin{equation}\label{eq:mir-simp}
	\M = \sigma_z\nu_x,
\end{equation}
where the factor of $i$ has been dropped. To proceed, we introduce the following unitary rotation to simultaneously diagonalize the chiral and mirror operators. 
\begin{equation}\label{eq:unitary}
		U(k_c) = \exp\left(+i\tfrac{k_c}{4}\nu_z\right)\exp\left(+i\tfrac{\pi}{4}\tau_z\right)
\end{equation}
After rotating the basis with $U$, the chiral operator becomes $\mathcal{C} = \sigma_z\nu_x\tau_z = \M\tau_z$, and the rotated Hamiltonian on the $k_c = \pi$ plane is
\begin{align}
\begin{split}
	\H(k_a,k_b,\pi) &= \Re{\E^p_{\k}}\tau_y - \Im{\E^p_{\k}}\sigma_z\tau_x + \\
&\hspace{5mm}-[\Re{\E^{po}_{\k}}\sigma_y + \Im{\E^{po}_{\k}}\sigma_x]\nu_z\tau_x \\
&\hspace{5mm}-[\Re{\E^{do}_{\k}}\sigma_y + \Im{\E^{do}_{\k}}\sigma_x]\nu_y\tau_x.
\end{split}
\end{align}
For simplicity we have neglected $\E_{\k}^d$, but it will be included in the final result. 

Translational symmetry is broken in the $\hat{\bm{b}}$ direction so $k_b$ is no longer a good quantum number. In the spirit of \citet{jackiw1976solitons} we expand the above Hamiltonian around $k_b = \pi$ (U-R-S-X plane containing the nodal ring) by introducing $p_b = k_b - \pi$ which we substitute as a real space derivative operator $-i\partial_b$ . Performing this expansion to linear order in $p_b$ we obtain
\begin{align}
\begin{split}
\H(k_a,p_b) &= \bigr[-2t_p\cos(\tfrac{1}{2}k_a)\tau_y - t_p'\cos(\tfrac{1}{2}k_a)\sigma_z\tau_x \\
	&\hspace{5mm} + \tfrac{1}{2}(t_{1p}^p + t_{2p}^o)\cos(\tfrac{1}{2}k_a)(\sigma_y - \sigma_x)\nu_z\tau_x \\
	&\hspace{5mm} + \tfrac{1}{2}t_d^o\sin(\tfrac{1}{2}k_a)(\sigma_y + \sigma_x)\nu_y\tau_x\bigr]p_b \\
	&\hspace{2mm}- \bigr[(t_{1p}^o - t_{2p}^o)\sin(\tfrac{1}{2}k_a)(\sigma_y + \sigma_x)\nu_z\tau_x \\
	&\hspace{5mm} + t_d^o\cos(\tfrac{1}{2}k_a)(\sigma_y - \sigma_x)\nu_y\tau_x\bigr].
\end{split}
\end{align}

Next, we block-diagonalize in mirror even and odd subspaces to reduce the above $8\times 8$ model into $4\times 4$ blocks. This is done by introducing bonding and anti-bonding combinations which are eigenstates of $\nu_x$
\begin{equation}\label{eq:ba-states}
	\ket{b} = \frac{1}{\sqrt{2}}(\ket{B} + \ket{T}), \qquad \ket{a} = \frac{1}{\sqrt{2}}(\ket{B} - \ket{T}),
\end{equation} 
where $\ket{B}$ is localized to the bottom layer (composed of $B,R$ iridiums) and $\ket{T}$ to the top layer (composed of $Y,G$ iridiums). It is then clear that $\{\ket{b\uparrow},\ket{a\downarrow}\}$ and $\{\ket{a\uparrow},\ket{b\downarrow}\}$ form bases for the four dimensional $\M = \pm 1$ subspaces, respectively, and the chiral operator reduces to $\mathcal{C} = \tau_z$. We introduce a new set of Pauli matrices $\bm{\eta}$ which act on the appropriate mirror subspace through
\begin{align}
\begin{split}
	\M = +: &\qquad \ket{b\uparrow}\ \stackrel{\eta_x}{\longleftrightarrow}\ \ket{a\downarrow}, \\
	\M = -: &\qquad \ket{a\uparrow}\ \stackrel{\eta_x}{\longleftrightarrow}\ \ket{b\downarrow}.
\end{split}
\end{align}
Within $\M = \pm$ the Hamiltonian becomes
\begin{align}
\begin{split}
	\H^{\pm}(k_a,p_b) &= \bigr[-2t_1(k_a)\id\tau_y - t_2(k_a)\eta_z\tau_x \\
	&\hspace{5mm}	+ t_{3\pm}(k_a)(\eta_y - \eta_x)\tau_x]p_b \\
	&\hspace{5mm} - t_{4\pm}(k_a)(\eta_y + \eta_x)\tau_x,
\end{split}
\end{align}
where
\begin{align}\label{eq:t-functions}
\begin{split}
	t_1(k_a) &= t_p\cos(\tfrac{1}{2}k_a), \\
	t_2(k_a) &= t_p'\cos(\tfrac{1}{2}k_a), \\
	t_{3\pm}(k_a) &= \tfrac{1}{2}(t_{1p}^o + t_{2p}^o)\cos(\tfrac{1}{2}k_a) \mp \tfrac{1}{2}t_d^o\sin(\tfrac{1}{2}k_a), \\
	t_{4\pm}(k_a) &= (t_{1p}^o - t_{2p}^o)\sin(\tfrac{1}{2}k_a) \pm t_d^o\cos(\tfrac{1}{2}k_a).
\end{split}
\end{align}

We then solve the Schr\"odinger equation to obtain the wave function for the zero modes in each mirror subspace
\begin{equation}
	\H^{\pm}(k_a,p_b \mapsto -i\partial_b)\Psi_{\pm}(k_a,x_b) = 0\cdot\Psi_{\pm}(k_a,x_b),
\end{equation}
with the wave function ansatz 
\begin{equation}\label{eq:wf-ansatz}
	\Psi_{\pm}(k_a,x_b) \propto e^{\lambda^{\pm}x_b}(\eta_x + \eta_y)\tau_x\chi_{\pm},
\end{equation}
where the factor $(\eta_x + \eta_y)\tau_x$ is chosen to simplify the following equations. The eigenequation for the four-component vector $\chi_{\pm}$ is
\begin{equation}\label{eq:pre-eigenequation}
	\{[2t_1(\eta_x + \eta_y)\tau_z + t_2(\eta_x - \eta_y) - 2t_{3\pm}\eta_z]\lambda^{\pm} - 2t_{4\pm}\}\chi_{\pm } = 0.
\end{equation}
Since we have a chiral symmetry $\mathcal{C} = \tau_z$ at $k_c = \pi$, this can be further reduced into a $2\times 2$ problem by block-diagonalizing in the appropriate chiral subspace. The outermost iridiums on this surface are $R,G$, so we diagonalize in $\tau_z = -1$. If we had instead considered the half space $x_b > 0$, the outermost iridiums would be $B,Y$ and we would diagonalize in $\tau_z = +1$. As the wave function ansatz contains $\tau_x$, which flips the eigenvalue of $\tau_z$, we instead set $\tau_z = +1$ which yields
\begin{equation}\label{eq:eigenequation}
	\{[(2t_1 + t_2)\eta_x + (2t_1 - t_2)\eta_y - 2t_{3\pm}\eta_z]\lambda^{\pm} - 2t_{4\pm}\id\}\chi_{\pm}^{BY} = 0.
\end{equation}
This describes a two-component vector $\chi_{\pm}^{BY}$ which must be an eigenstate of 
\begin{equation}\label{eq:22-hamiltonian}
	h^{\pm} = \bm{d_{\pm}}\cdot\bm{\eta}, \qquad \bm{d_{\pm}} = (2t_1 + t_2, 2t_1 - t_2, -2t_{3\pm}),
\end{equation}
with eigenvalue $(-1)^j\norm{\bm{d_{\pm}}}$, where $j = 0,1$. The eigenstates of $h^{\pm}$ can be parameterized with the angles defined through
\begin{equation}\label{eq:parameterization}
\frac{\bm{d_{\pm}}}{\norm{\bm{d_{\pm}}}} = (\sin\theta_{\pm}\cos\varphi_{\pm},\sin\theta_{\pm}\sin\varphi_{\pm},\cos\theta_{\pm}).
\end{equation}
Solutions $\chi_{\pm,j}^{BY}$ corresponding to eigenvalue $(-1)^j\norm{\bm{d_{\pm}}}$ in the mirror even subspace are
\begin{align}
\begin{split}\label{eq:mp-chi}
	\chi_{+,0}^{BY} &= \cos(\tfrac{1}{2}\theta_+)\ket{b\uparrow} + e^{i\vp_+}\sin(\tfrac{1}{2}\theta_+)\ket{a\downarrow},\\
	\chi_{+,1}^{BY} &= \sin(\tfrac{1}{2}\theta_+)\ket{b\uparrow} - e^{i\vp_+} \cos(\tfrac{1}{2}\theta_+)\ket{a\downarrow},
\end{split}
\end{align}
while those in the mirror odd subspace are
\begin{align}
\begin{split}\label{eq:mm-chi}
	\chi_{-,0}^{BY} &= \cos(\tfrac{1}{2}\theta_-)\ket{a\uparrow} + e^{i\vp_-}\sin(\tfrac{1}{2}\theta_-)\ket{b\downarrow} \\
	\chi_{-,1}^{BY} &= \sin(\tfrac{1}{2}\theta_-)\ket{a\uparrow} - e^{i\vp_-} \cos(\tfrac{1}{2}\theta_-)\ket{b\downarrow}.
\end{split}
\end{align}

\begin{figure}
	\includegraphics[width=\linewidth]{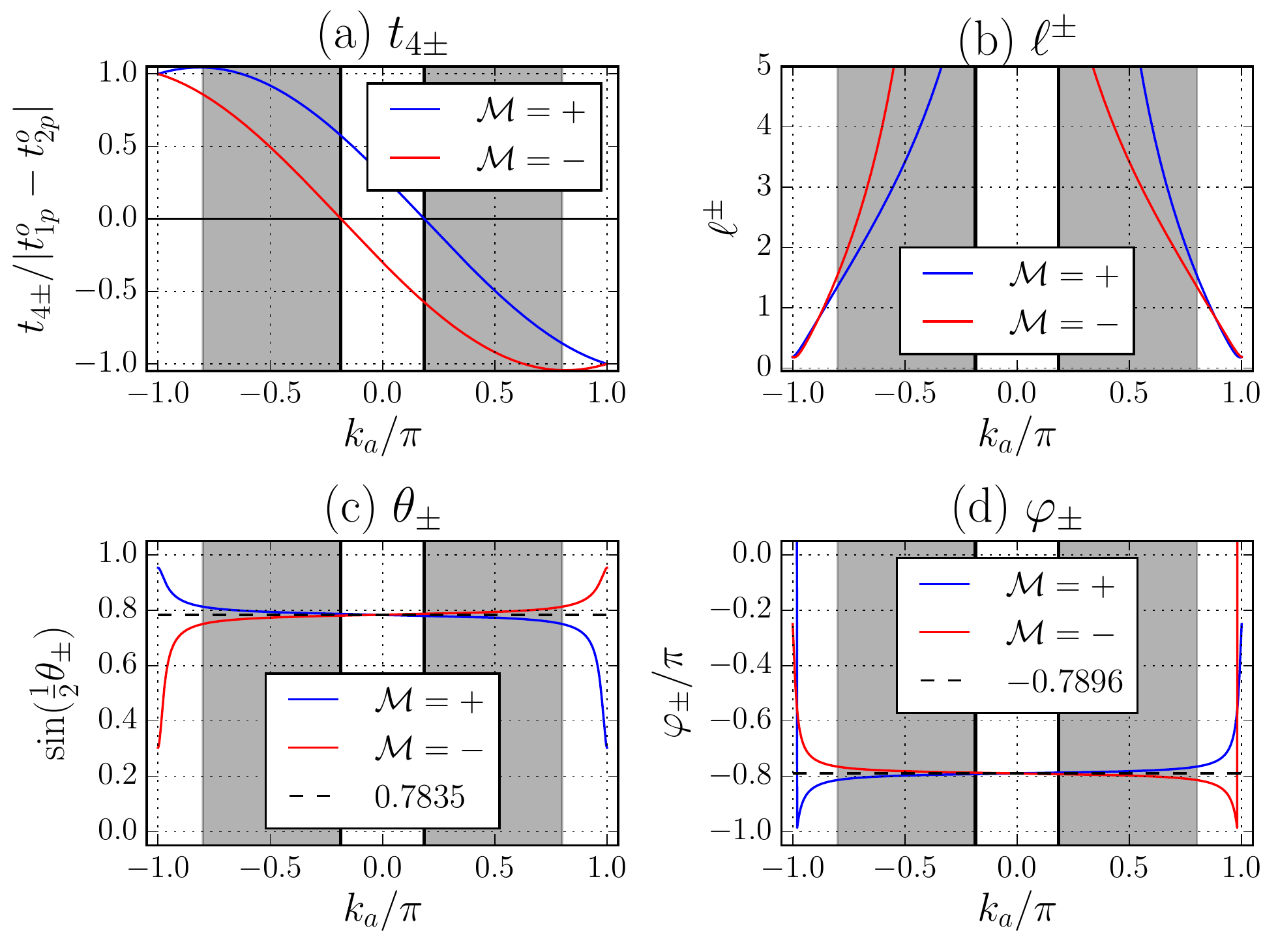}
	\caption{(colour online) Thick vertical lines indicate the position of the nodal ring, and the shaded area corresponds to the $k_a$ for which our model remains valid. (a) Value of $t_{4\pm}$ as a function of $k_a$ whose sign determines the solution. (b) Penetration depth $\ell^{\pm}$ in units of the $\hat{\bm{b}}$ lattice spacing for the physically relevant solution in each region outside the nodal ring. (c) $\sin(\tfrac{1}{2}\theta_{\pm})$ and (d) $\vp_{\pm}$ parameters. The number of digits quoted reflects the agreement with angles averaged over $k_c$ obtained from the wave functions calculated in a finite slab geometry. All quantities are calculated with the tight-binding parameters for SrIrO$_3$ given in Table~\ref{table:tb-parameters}.}
	\label{fig:wf-parameters}
\end{figure}

Substituting these solutions into the eigenequation Eq.~\ref{eq:eigenequation} we find the exponential decay parameter $\lambda^{\pm}_j$
\begin{equation}\label{eq:decay-constant}
\lambda^{\pm}_j(k_a) = \frac{2(-1)^jt_{4\pm}(k_a)}{\norm{\bm{d_{\pm}}(k_a)}}.
\end{equation}
For each $k_a$ the physically relevant solution will be $\chi_{\pm,j}^{BY}$ with $\lambda_j^{\pm} > 0$ because the wave function should decay into the bulk as $x_b \rightarrow -\infty$. The value of $j$ is determined by the sign of $t_{4\pm}$ which is plotted in Fig.~\ref{fig:wf-parameters}(a). For $k_a < 0$, $t_{4\pm} > 0$ and the solutions are $\chi_{\pm,0}^{BY}$, while for $k_a > 0$, $t_{4\pm} < 0$ so the solutions are $\chi_{\pm,1}^{BY}$. The exponential decay parameters $\lambda_1^{\pm}(+k_a)$ are equal to $\lambda_0^{\mp}(-k_a)$, so we will simply write $\lambda_j^{\pm} = \lambda^{\pm}$. The decay parameter sets the penetration depth $\ell^{\pm} = 1/\lambda^{\pm}$ of the wave functions into the bulk, which is plotted in Fig.~\ref{fig:wf-parameters}(b). To obtain the final form of the wave function we must perform the rotation $(\eta_x + \eta_y)\tau_x$ appearing in the wave function ansatz Eq.~\ref{eq:wf-ansatz}. The first $\tau_x$ operation simply changes the composition from sublattice $B,Y$ to $R,G$. For $k_a < 0$, $\chi_{+,0}^{RG}$ is rotated to
\begin{align}
	\begin{split}
	&\sim (1 - i)\sin(\tfrac{1}{2}\theta_+)\ket{b\uparrow} + (1 + i)e^{-i\vp_+}\cos(\tfrac{1}{2}\theta_+)\ket{a\downarrow} \\
	&\sim \sin(\tfrac{1}{2}\theta_+)\ket{b\uparrow} + e^{+i(\tfrac{\pi}{2} - \vp_+)}\cos(\tfrac{1}{2}\theta_+)\ket{a\downarrow},
	\end{split}
\end{align}
with a similar result for $\chi_{-,0}^{RG}$. For $k_a > 0$, $\chi_{+,1}^{RG}$ is rotated to
\begin{align}
	\begin{split}
	&\sim (1 - i)\cos(\tfrac{1}{2}\theta_+)\ket{b\uparrow} - (1 + i)e^{-i\vp_+}\sin(\tfrac{1}{2}\theta_+)\ket{a\downarrow} \\
	&\sim \cos(\tfrac{1}{2}\theta_+)\ket{b\uparrow} + e^{-i(\tfrac{\pi}{2} + \vp_+)}\sin(\tfrac{1}{2}\theta_+)\ket{a\downarrow},
	\end{split}
\end{align}
with a similar result for $\chi_{-,1}^{RG}$.

Close to $k_a = 0$, $t_{4\pm}$ changes sign when
\begin{equation}\label{eq:nr-boundary}
	|k_a^*| \approx 2\tan(\tfrac{1}{2}|k_a^*|) =  \frac{2t_d^o}{|t_{1p}^o - t_{2p}^o|},
\end{equation}
which corresponds to the semi-major axis of the bulk nodal ring ellipse\cite{rhim2015landau}. Across the nodal ring one of $\lambda^{\pm}$ vanishes, signaling a sharp change in the wave function due to closing of the bulk gap. A similar sharp change must happen across the zone boundary $k_a = \pi$ which separates $k_a > 0$ and $k_a < 0$ solutions. We therefore focus on $k_a$ outside the nodal ring and away from the zone boundary; the shaded region in Fig.~\ref{fig:wf-parameters}. 

Therefore, for $k_a < 0$ the wave functions describing the zero modes are
\begin{align}
\begin{split}\label{eq:wf-ka-p}
	\Psi_+ &= e^{\lambda^+x_b}[\cos(\tfrac{1}{2}\theta_+)\ket{b\uparrow} + e^{-i(\tfrac{\pi}{2} + \vp_+)}\sin(\tfrac{1}{2}\theta_+)\ket{a\downarrow}] \\
	\Psi_- &= e^{\lambda^-x_b}[\cos(\tfrac{1}{2}\theta_-)\ket{a\uparrow} + e^{-i(\tfrac{\pi}{2} + \vp_-)}\sin(\tfrac{1}{2}\theta_-)\ket{b\downarrow}],
\end{split}
\end{align}
where $\lambda^{\pm}$, $\theta_{\pm}$, and $\vp_{\pm}$ are all functions of $k_a$ and plotted in Fig.~\ref{fig:wf-parameters}(b-d). Solutions for $k_a < 0$ are related to those above by time-reversal symmetry, for $\Psi_{\pm}(k_a < 0)$ is the time-reversal partner of $\Psi_{\mp}(k_a > 0)$. The states $\ket{B},\ket{T}$ appearing in $\ket{b},\ket{a}$ (Eq.~\ref{eq:ba-states}) are understood to be iridium $R,G$ states, respectively. In the region of interest away from the nodal ring and zone boundary, to good approximation the angles $\theta_{\pm},\vp_{\pm}$ can be taken as constants $\theta,\vp$ with values given in Fig.~\ref{fig:wf-parameters}(c-d). This is because away from $k_a = \pi$, the relevant functions in Eq.~\ref{eq:t-functions} are all dominated by $\cos(\tfrac{1}{2}k_a)$. Owing to the small size of $t_d^0, t_d'$ due to slight rotation and tilting of oxygen octahedra, the cosine varies slowly to yield nearly constant angles defined through Eq.~\ref{eq:parameterization}. We find that $\theta_{\pm}(k_a),\varphi_{\pm}(k_a)$ remain within 10\% of their value at $k_a = 0$ up to $|k_a| \approx 0.88\pi$ and $|k_a| \approx 0.93\pi$, respectively.

This result also holds for an $\hat{\bm{a}}$ termination in the half space $x_a > 0$ if we simply switch the labels $a\leftrightarrow b$. The solution for $x_b > 0$ ($x_a < 0$) will involve diagonalization in the  $\tau_z = +1$ subspace, since the outermost iridiums are $B,Y$, and the solutions for $k_a < 0$ and $k_a > 0$ will be switched with $\ket{B},\ket{T}$ as $B,Y$ iridium states. 

As a final note, the small $\E_{\k}^d$ term we have ignored contributes 
\begin{equation}
	\Im{\E_{\k}^d}\nu_x\tau_x = -t_d'\sin(\tfrac{1}{2}k_a)p_b\nu_x\tau_x,
\end{equation}
which becomes
\begin{equation}
	\mp t_d'\sin(\tfrac{1}{2}k_a)p_b\eta_z\tau_x 	
\end{equation}
in the mirror even and odd subspaces. This serves to modify the function $t_2$
\begin{equation}
	t_2(k_a) \mapsto t_{2\pm}(k_a) = t_p'\cos(\tfrac{1}{2}k_a) \pm t_d'\sin(\tfrac{1}{2}k_a),
\end{equation}
which slightly changes the wave function parameters $\lambda^{\pm}, \theta_{\pm},\vp_{\pm}$, but cannot change the form of the solutions. 

\subsection{\label{subsec:slab-calculations} Extending Solution with Slab Calculations}

\begin{figure}
	\includegraphics[width=0.7\linewidth]{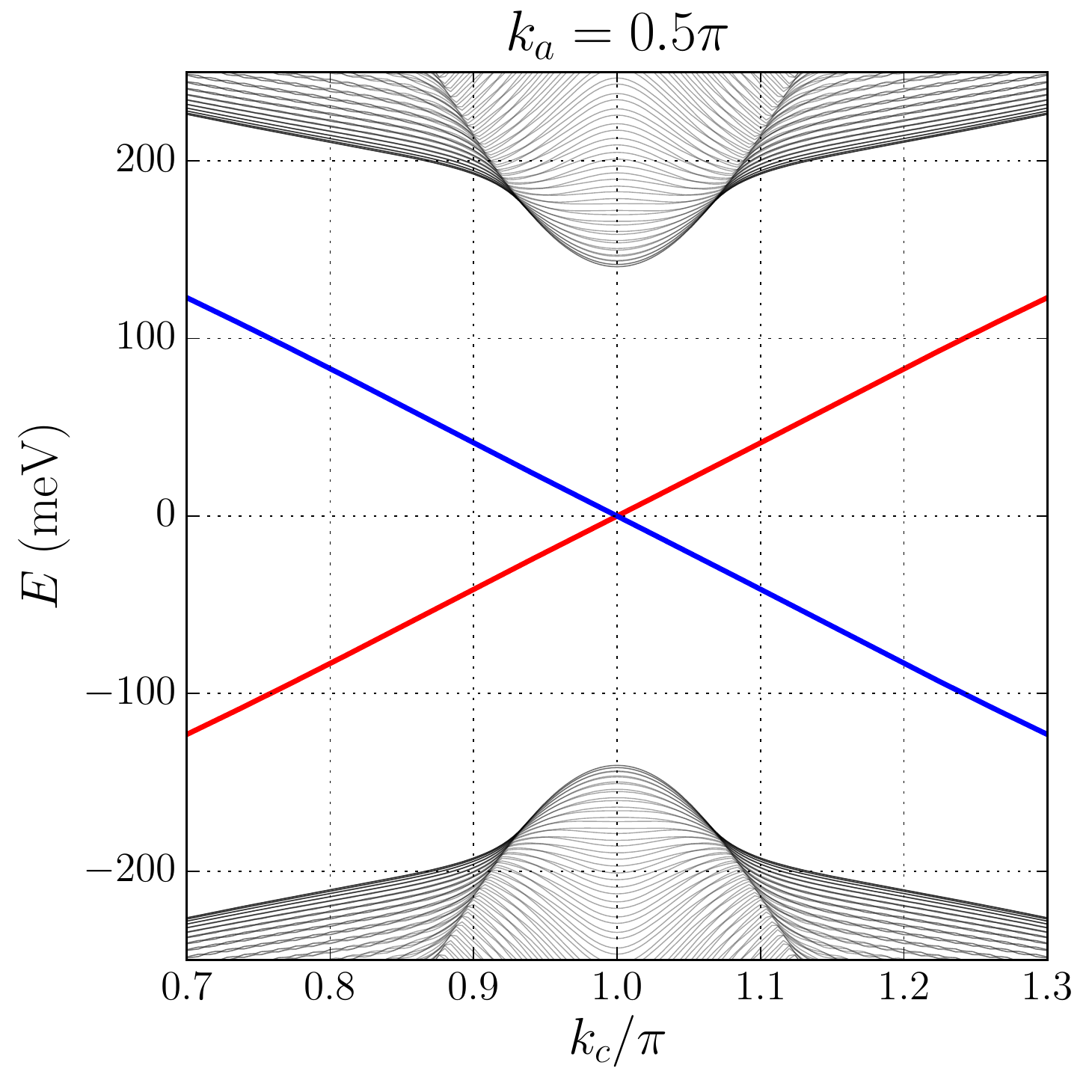}
	\caption{Dispersion of surface zero modes away from $k_c = \pi$ at $k_a = 0.5\pi$, obtained from exact diagonalization of the tight-binding model in a $\hat{\bm{b}}$ slab geometry with $N = 250$ layers. Blue and red colours represent mirror eigenvalues $\M = \pm 1$, respectively.}
	\label{fig:slab-dispersion}
\end{figure}

To extend the analytic solution for $\Psi_{\pm}$ away from $k_c = \pi$, we numerically diagonalize the tight-binding Hamiltonian Eq.~\ref{eq:full-tb} in a $\hat{\bm{b}}$ slab geometry to obtain the wave functions localized to the appropriate edge. At $k_c = \pi$, we find $\braket{\tau_z} = -1$ at the edge with outer $R,G$ iridiums, which is the appropriate chiral subspace for that surface. For $k_c \ne \pi$ the Hamiltonian no longer has a chiral symmetry, and we find $\braket{\tau_z} \ne -1$ due to mixing with the other chiral subspace, $\tau_z = +1$. However, this deviation does not affect the main conclusion as discussed below.

We find that the expectation of the mirror operator $\braket{\M} = \braket{\sigma_z\nu_x}$ in the two branches remains $\pm 1$ away from $k_c = \pi$. For $k_a < 0$ the left-moving branch has $\braket{\M} = -1$ and the right $\braket{\M} = +1$, with the opposite holding for $k_a > 0$ as shown in Fig.~\ref{fig:slab-dispersion} for $k_a = 0.5\pi$. Examining the wave function components, they are found to be the appropriate combination of bonding and anti-bonding states. Finally, the weights of $\ket{b\sigma}$ and $\ket{a\sigma}$ for $k_c \ne \pi$ are found to be in excellent agreement with the analytic expressions Eq.~\ref{eq:wf-ka-p} at $k_c = \pi$, which are plotted in Fig.~\ref{fig:wf-parameters}. In particular, for $|k_c - \pi| \le 0.3\pi$ the angles $\theta_{\pm},\varphi_{\pm}$ were found to remain within 5\%, 15\% of their value at $k_c = \pi$, respectively, in the $k_a$ region of validity discussed below.

Therefore, our numerical results imply that the wave functions in Eq.~\ref{eq:wf-ka-p} describe mirror even and odd branches away from $k_c = \pi$. The dispersion of the branches with mirror eigenvalue $m = \pm$ are
\begin{equation}\label{eq:ss-dispersion}
	\varepsilon_{\k m} = m\cdot v_F\cdot  \mathrm{sgn}(-k_a)\cdot(k_c - \pi),
\end{equation}
where $v_F$ is the velocity of the surface states. From the slab calculation we estimate this velocity to be $v_F \approx (c/\mathrm{\AA})(2.0 \times 10^4\ \mathrm{m/s}),$ where $c$ is the $\hat{\bm{c}}$ lattice spacing. For SrIrO$_3$ grown on a SrTiO$_3$ substrate, $c \approx 7.97\ \mathrm{\AA}$\cite{kim2015surface} which yields $v_F \approx 1.6\times 10^5\ \mathrm{m/s}$.

To determine the values of momentum for which we have localized surface states, one must consider the proximity of bulk states. The exponential decay parameter given by Eq.~\ref{eq:decay-constant} determines the penetration depth $\ell^{\pm} = 1/\lambda^{\pm}$ of the wave function into the bulk, which should be compared to the thickness $L$ of the finite slab. In our numerical study we consider $L = 250$ in units of the $\hat{\bm{b}}$ lattice spacing, which satisfies $L \gg \ell^{\pm}(k_a)$ for $k_a$ away from the nodal ring. As $|k_a|$ approaches $|k_a^*|$, the penetration depth becomes large due to mixing with extended bulk states. For each $k_a$ the minimum of the bulk bands determines the high-energy (and momentum) cutoff $\Lambda(k_a)$ below which the low energy theory is valid. The cutoff vanishes at the nodal ring $|k_a^*|$ and becomes small near the zone boundary $|k_a| = \pi$. In order to have a constant cutoff $\Lambda_0$, the $k_a$ region is defined to be those for which the cutoff is at least $\Lambda_0$: $\{k_a\ |\ \Lambda(k_a) \ge \Lambda_0\}$. This constrains $k_c$ to be in $[-\Lambda_0/v_F,+\Lambda_0/v_F]$, which determines the range of momentum used in the electron-phonon interaction, shown as the $x$-axis in Fig.~\ref{fig:ph-excitations}(a-b). With tight binding parameters for SrIrO$_3$, and $\Lambda_0 \approx 100\ \mathrm{meV}$ close to its maximum value, the $|k_a|$ region is approximately  $ [0.4\pi,0.75\pi]$. 
However, in the real material this region is expected to be larger due to the small size of the nodal ring. Note that the main conclusion, i.e. qualitative difference in the damping of mirror odd and even phonon modes, is independent of this cutoff. 

\subsection{\label{subsec:densities} Total and Relative Density}
The wave functions obtained for $k_a > 0$ and $k_a < 0$ in each mirror subspace Eq.~\ref{eq:mp-chi},~\ref{eq:mm-chi} are linearly independent solutions of the same $2\times 2$ Hamiltonian Eq.~\ref{eq:22-hamiltonian}. This means the wave function expressions in Eq.~\ref{eq:wf-ka-p} can be inverted to write bonding and anti-bonding states in terms of the surface wave functions. It is easily shown that
\begin{align}
\begin{split}
	b_{\uparrow}^{\dag}b_{\uparrow} + a_{\downarrow}^{\dag}a_{\downarrow} &= e^{-2\lambda^+x_b}\Psi_+^{\dag}(+k_a)\Psi_+(+k_a) + (-k_a) \\
	a_{\uparrow}^{\dag}a_{\uparrow} + b_{\downarrow}^{\dag}b_{\downarrow} &= e^{-2\lambda^-x_b}\Psi_-^{\dag}(+k_a)\Psi_-(+k_a) + (-k_a),
\end{split}
\end{align} 
where $b_{\sigma},a_{\sigma}$ are electron operators for the states $\ket{b\sigma},\ket{a\sigma}$, and $\Psi_{\pm}$ for the surface states. In the above $k_a$ is assumed to be positive, and $(-k_a)$ represents the contribution from the $-k_a$ region. Therefore the total density
\begin{equation}
	\rho_+ = \sum_{\sigma}(b_{\sigma}^{\dag}b_{\sigma} + a_{\sigma}^{\dag}a_{\sigma})
\end{equation}
can be written in terms of $\Psi_{\pm}^{\dag}\Psi_{\pm}$ connecting states with the same mirror eigenvalue. This can be simplified close to $k_c = \pi$
\begin{equation}
	\rho_+ = \sum_{\sigma}(c_{R\sigma}^{\dag}c_{R\sigma}^{\phantom{\dag}} + c_{G\sigma}^{\dag}c_{G\sigma}^{\phantom{\dag}}),\nonumber
\end{equation}
where the wave function is dominantly supported on $R,G$ iridiums, and $\braket{\tau_z} \approx -1$. Clearly, the total density is even under mirror reflection. Similarly, with
\begin{align}
\begin{split}
	b_{\uparrow}^{\dag}a_{\uparrow} + a_{\downarrow}^{\dag}b_{\downarrow} &= e^{-(\lambda^+ + \lambda^-)x_b}\Psi_+^{\dag}(+k_a)\Psi_-(+k_a) + (-k_a) \\
	a_{\uparrow}^{\dag}b_{\uparrow} + b_{\downarrow}^{\dag}a_{\downarrow} &= e^{-(\lambda^+ + \lambda^-)x_b}\Psi_-^{\dag}(+k_a)\Psi_+(+k_a) + (-k_a)
\end{split}
\end{align}
the relative density between different layers
\begin{equation}
	\rho_- = \sum_{\sigma}(b_{\sigma}^{\dag}a_{\sigma} + a_{\sigma}^{\dag}b_{\sigma})
\end{equation}
can be written in terms of $\Psi_{\pm}^{\dag}\Psi_{\mp}$ connecting states with different mirror eigenvalues, and is therefore odd under mirror reflection. It can also be simplified as
\begin{equation}
	\rho_- = \sum_{\sigma}(c_{R\sigma}^{\dag}c_{R\sigma}^{\phantom{\dag}} - c_{G\sigma}^{\dag}c_{G\sigma}^{\phantom{\dag}}), \nonumber
\end{equation}
close to $k_c = \pi$. The momenta $k',k$ appearing in $\Psi^{\dag}(k')\Psi(k)$ need not be same because the angles $\theta,\vp$ in the wave function are approximately constant. So long as $k',k$ lie in the same $k_a$ region with small difference in $k_c$, the above relations hold. As shown in the next section, mirror even phonons can only couple to the total density while mirror odd phonons can only couple to the relative density. 
 
\section{\label{sec:epi-polarization} Electron-Phonon Coupling and Density Response}

In this section we focus on optical phonons in \textit{Pbnm}-AIrO$_3$ with momentum $\q$ close to the zone center. Before considering the effect of terminating the crystal, we review the symmetry of bulk optical phonons. The zone center phonons of the isostructural \textit{Pbnm}-SrHfO$_3$ have been classified according to irreducible representations of the point group $D_{2h}$\cite{fatteley1972infrared, park1976raman, vali2009lattice}
\begin{align}
\begin{split}
&\Gamma_O: \resizebox{220pt}{!}{$7A_g\oplus 5B_{1g}\oplus 7B_{2g}\oplus 5B_{3g}\oplus 8A_u\oplus 9B_{1u}\oplus 7B_{2u}\oplus 9B_{3u}$} \\
&\Gamma_A: \resizebox{67.75pt}{!}{$B_{1u}\oplus B_{2u}\oplus B_{3u}$},
\end{split}
\end{align}
where $\Gamma_O,\Gamma_A$ refer to optical and acoustic modes, respectively. Of the optical modes, 25 are infrared (IR) active ($B_{1u},B_{2u},B_{3u}$) and 24 are Raman active ($A_g, B_{1g},B_{2g},B_{3g}$). The IR active modes $B_{1u},B_{3u},B_{2u}$ are polarized along the $\hat{\bm{a}}, \hat{\bm{b}}, \hat{\bm{c}}$ directions, respectively.

\begin{table}
\caption{\label{table:c1h-chartable} Character table for the point group $C_{1h}\ (m)$.\cite{dresselhaus}}
\begin{tabular}{c | c | c c}
\toprule
\multicolumn{2}{c|}{$C_{1h}\ (m)$}& $E$ & $\sigma_c$ \\
\hline 
$a,b$ & $A'$ & $1$ & $+1$ \\
$c$ & $A''$ & $1$ & $-1$  \\
\toprule
\end{tabular}
\end{table}

\begin{figure}
	\includegraphics[width=0.71\linewidth]{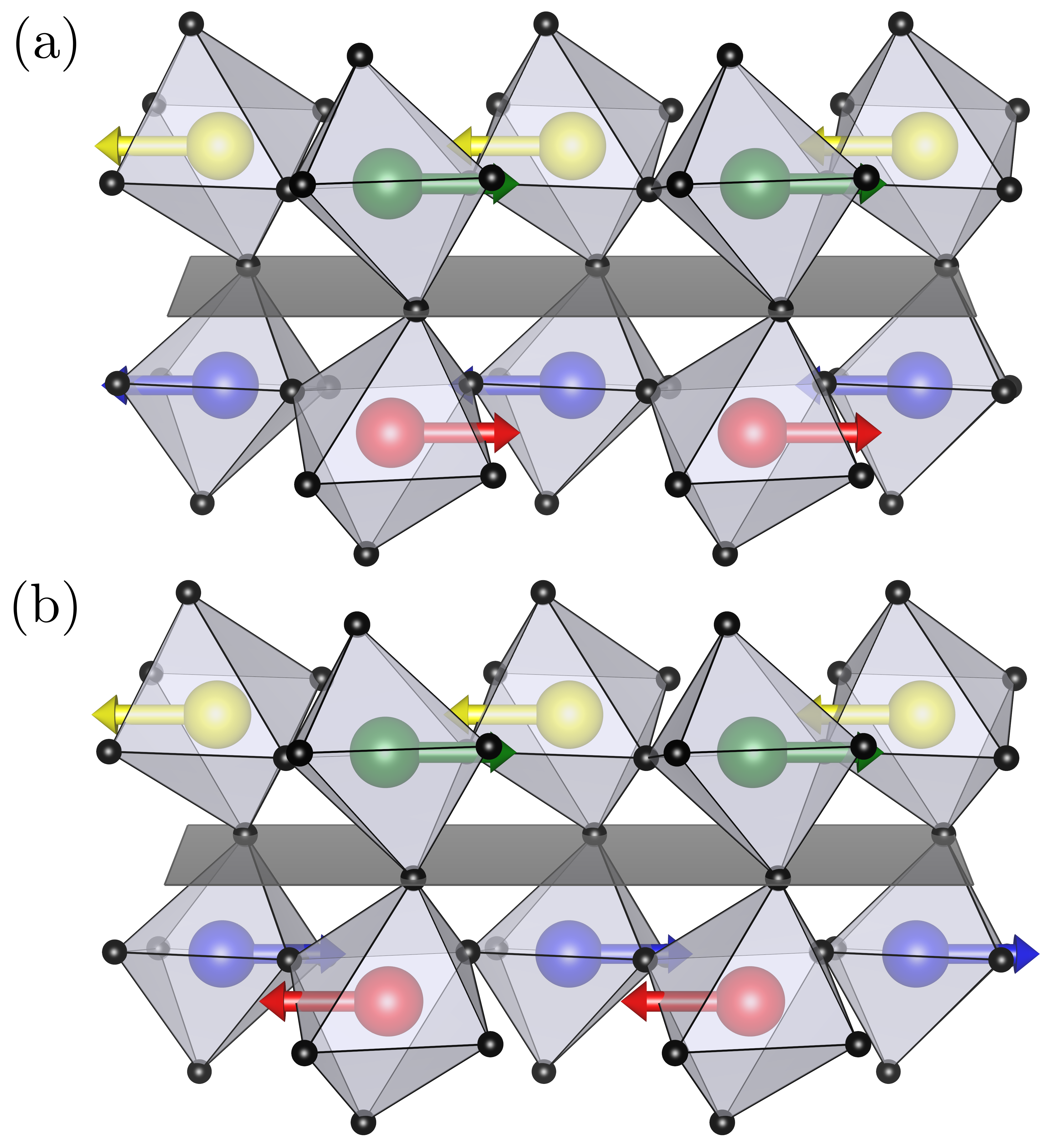}
	\caption{(colour online) Schematic long-wavelength mirror (a) even and (b) odd optical phonon modes on a $\hat{\bm{b}}$ surface. Arrows represent the relative displacement between iridium and oxygen atoms.}
	\label{fig:phonons} 
\end{figure}

For a crystal terminated in the $\hat{\bm{b}}$ direction with $N$ layers, we perform a straightforward classification of the zone center phonons. For simplicity we focus on modes involving displacements between iridium and oxygen atoms. In this slab configuration only the mirror symmetry remains, which leads to the point group $C_{1h}\ (m)$ with character table shown in Table~\ref{table:c1h-chartable}. Following the notation of \citet{dresselhaus}, the vector representation is given by 
\begin{equation}
	\Gamma_{\mathrm{vec}} = 2A'\oplus A''.	
\end{equation}
Under mirror reflection all the iridiums are switched ($B\leftrightarrow Y$, $R\leftrightarrow G$), so the character of the mirror operation in the equivalence representation is zero. The character of the identity operation is simply $4N$, so the equivalence representation can be easily decomposed as
\begin{equation}
	\Gamma_{\mathrm{equiv}} = 2NA'\oplus 2NA''.
\end{equation}
Therefore, at the zone center the representation of these phonon modes is
\begin{equation}
	\Gamma_{\mathrm{phon}} = \Gamma_{\mathrm{vec}}\otimes \Gamma_{\mathrm{equiv}} = 6NA'\oplus 6NA''.
\end{equation}
Each mode has a simple interpretation because $\chi_{A'}(\sigma_c) = +1$ and $\chi_{A''}(\sigma_c) = -1$; they have definite parity under mirror reflection. This analysis holds for finite $q_a$ as the mirror symmetry is preserved. In terms of atomic displacements, the meaning of mirror even and odd phonon modes is that the displacement $\bm{\xi}^{\alpha}$ of ion $\alpha$ is mapped to $\pm \bm{\xi}^{\alpha'}$ of ion $\alpha'$ under reflection. This is illustrated in Fig.~\ref{fig:phonons} for modes involving relative Ir and O displacements. 

Restricting ourselves to optical phonons of definite mirror symmetry, we will broadly label the modes by $\lambda = \pm$ with field operator $A_{\q\lambda} = a_{\q\lambda} + a_{-\q\lambda}^{\dag}$, displacements $\bm{\xi}_{\q\lambda}^{\alpha}$, and unperturbed Matsubara Green's function\cite{mahan}
\begin{equation}\label{eq:unperturbed-propagator}
	\mathcal{D}_{\lambda}^0(\q,iq_n) = \frac{2\omega_{\q\lambda}^0}{(iq_n)^2 - (\omega_{\q\lambda}^0)^2}, \qquad q_n = \frac{2\pi n}{\beta},
\end{equation}
where $\omega_{\q\lambda}^0$ is the dispersion, and $q_n$ the Matsubara frequency. 

The aim of this section is to show how certain optical phonons with $\q$ along $q_a$ close to the zone center are damped through their interaction with electronic surface states. First, we examine how mirror even and odd modes couple to the surface electrons through microscopic and symmetry considerations. We then discuss a certain type of surface localized phonon, and how it couples differently to electrons than bulk phonons. Finally, we calculate the imaginary part of the first-order phonon self-energy $\Pi_{\lambda}^0$.

\subsection{\label{subsec:ep-vertices} Symmetry-Allowed Electron-Phonon Vertices}

For electrons tightly-bound to Ir sites, their interaction with bulk phonon modes involving Ir displacements takes the general form\cite{mahan}
\begin{equation}\label{eq:frohlich-coupling}
	\mathcal{H}_{ep} = 	\sum_{\q\lambda\alpha}\underbrace{\left(-i\sqrt{\frac{\hbar}{2M_{\alpha}\omega_{\q\lambda}^0}}(\hat{\bm{\xi}}^{\alpha}_{q\lambda}\cdot\q)V_{\alpha}(\q)\right)}_{g_{\q\lambda\alpha}}A_{\q\lambda}\rho_{\q\alpha},
\end{equation}
where $\alpha\in\{B,R,Y,G\}$, $V_{\alpha}$ is the atomic potential, and $\rho_{\alpha}$ is the density of electrons on Ir site $\alpha$. Longitudinal optical phonons have displacements proportional to $\hat{\q}$ and nearly flat dispersion near the zone center, so the electron-phonon coupling $g_{\q\lambda\alpha}$ scales like
\begin{equation}
	g_{\q\lambda\alpha} \propto (\hat{\q}\cdot\q)\frac{1}{q^2} = \frac{1}{q},
\end{equation}
which is the well-known \citet{frohlich1954} polar coupling in $d = 3$ dimensions with long-range Coulomb potential $V_{\alpha}(\q) \propto q^{-2}$.

As shown in Section~\ref{sec:surface-states}, the wave function describing surface electrons is dominantly supported on the $R,G$ iridiums near $k_c = \pi$, so for simplicity we neglect $B,Y$ sites. Along $q_a$ near the zone center, the phonons have definite mirror symmetry, and the displacements of the iridiums satisfy $\bm{\xi}_{\pm}^{R} = \pm \bm{\xi}^{G}_{\pm}$. This means the coupling constants satisfy $g_{\pm R} = \pm g_{\pm G}$. Taking their common value $g_{\pm}$, the surface electron-phonon interaction takes the form 
\begin{align}\label{eq:micro-ep-coupling}
\begin{split}
	\mathcal{H}_{ep} &= \sum_{\lambda = \pm }g_{\lambda}A_{\lambda}\sum_{\sigma}(c_{R\sigma}^{\dag}c_{R\sigma}^{\phantom{\dag}} + \lambda c_{G\sigma}^{\dag}c_{G\sigma}^{\phantom{\dag}} ) \\
					 &= \sum_{\lambda = \pm }g_{\lambda}A_{\lambda}\rho_{\lambda},
\end{split}
\end{align}
in which the total and relative densities of Section~\ref{subsec:densities} appear naturally. In a $\hat{\bm{b}}$ slab geometry, the electron-phonon coupling Eq.~\ref{eq:frohlich-coupling} must be modified as $q_b$ is no longer a good quantum number. However, properties of the displacements $\bm{\xi}^{\alpha}$ under mirror reflection lead to the same qualitative result. A modified electron-phonon coupling is discussed in the next section. 

\begin{figure}
	\includegraphics[width=\linewidth]{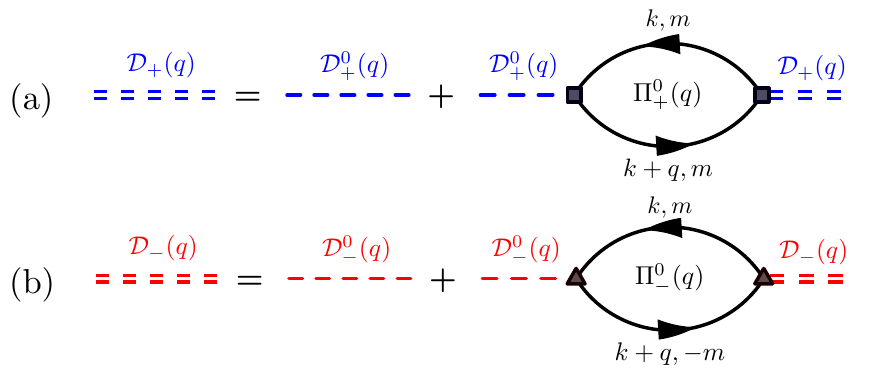}
	\caption{(colour online) Dyson equation for (a) even and (b) odd phonon propagators $\mathcal{D}_{\pm}(q)$ through their interaction with surface states for small $q_a$. The vertex $\Box$ represents even modes coupling through the total density, while $\triangle$ represents coupling  of odd modes through the relative density. As the irreducible self-energy we take the first-order particle-hole bubble $\Pi^0_{\pm}(q)$ in which we sum over $m$, indexing the mirror branch of the electrons.}
	\label{fig:phonon-RPA} 
\end{figure}

The form of the coupling Eq.~\ref{eq:micro-ep-coupling} can also be understood through symmetry considerations. Under mirror reflection the phonon operators transform as $\M A_{\pm}\M^{\dag} = \pm A_{\pm}$, and the densities as $\M\rho_{\pm}\M^{\dag} = \pm \rho_{\pm}$. Therefore, the only mirror-invariant electron-phonon vertices we can write are
\begin{equation}\label{eq:sym-ep-coupling}
	\H_{ep} \propto g_+A_+\rho_+ + g_-A_-\rho_-.
\end{equation}
Fig.~\ref{fig:phonon-RPA} shows Dyson's equation for the phonon propagators, where these vertices appear in the first-order self-energy. As shown in Section~\ref{subsec:densities}, the even modes $A_+$ coupling to the total density $\rho_+$ can only excite electron-hole pairs lying in the same mirror branch, while odd modes $A_-$ coupling to the relative density $\rho_-$ can excite pairs lying in different mirror branches. The form of this interaction is based on mirror symmetry alone, and is independent of the sublattice composition $\braket{\tau_z}$.

Note that form factor is independent of momentum, unlike carbon nanotubes\cite{ishikawa2006} where the low-energy Dirac physics leads to a spinor that varies dramatically with momentum. This is because the parameters $\theta_{\pm},\varphi_{\pm}$ describing the surface state wave functions remain nearly constant with momentum due to small rotation and tilting of oxygen octahedra. Consequently, the form factors remain nearly constant, with modulus close to unity.

\subsection{\label{subsec:fk-modes} Fuchs-Kliewer Modes}
When the crystal is terminated, generally there will be a large number of vibrational modes with frequencies lying between the bulk values with quantized wavelengths in the normal direction. Fuchs and Kliewer\cite{fuchs1965optical,kliewer1966optical1, kliewer1966optical2} studied long wavelength modes in a polar material, and found that optical modes localized to the surface can exist in addition to extended bulk-like modes. Out of phase motion between oppositely charged ions sets up a macroscopic polarization field, with associated electric and electric displacement fields, which are described by Maxwell's equations. By imposing boundary conditions for the field inside and outside the material, an exponentially localized electric field exists  provided that
\begin{equation}
	\frac{1}{\epsilon(\omega)}\left(q^2 - \epsilon(\omega)\frac{\omega^2}{c^2}\right)^{1/2} = -\left(q^2 - \frac{\omega^2}{c^2}\right)^{1/2},	
\end{equation}
where $q$ is the wavevector of the field parallel to the surface, $\omega$ is the frequency of the field, and $\epsilon(\omega)$ is the dielectric function of the material. A surface optical (SO) mode exists when $\epsilon(\omega) < 0$.  Using a simple independent oscillator model of the dielectric function
\begin{equation}
	\epsilon(\omega) = \epsilon_{\infty} + \frac{\epsilon_0 - \epsilon_{\infty}}{1 - (\omega/\omega_{TO})^2},	
\end{equation}
where $\omega_{TO}$ is the transverse optical (TO) phonon frequency, and $\epsilon_0,\epsilon_{\infty}$ are the low- and high-frequency dielectric constants, an SO mode exists when\cite{kliewer1966optical1,wang1972electron,devreese2013elementary}
\begin{equation}
	\omega_{TO} < \omega < \omega_{LO} \quad \mathrm{and} \quad q > \omega_{TO}/c.	
\end{equation}
Therefore, in a polar crystal with LO-TO splitting we expect an SO mode to exist between the IR active TO and LO mode frequencies at long wavelengths.

As with the bulk modes, the macroscopic polarization produced by the SO mode couples to the electric field of the electron through 
\begin{equation}
	\H_{ep} = \iint\d \bm{r}\d \bm{R}\ \Psi^{\dag}(\bm{r})\frac{e(\bm{r} - \bm{R})\cdot\bm{P}(\bm{R})}{\lVert \bm{r} - \bm{R}\rVert^3}\Psi(\bm{r}),
\end{equation}
where $\Psi(\bm{r})$ is the electron field operator, and $\bm{P}(\bm{R})$ is the polarization field. Due to the exponential attenuation of the mode amplitudes into the bulk, SO modes will couple differently to electrons than their bulk LO counterparts. Choosing the electron field to be the TCSM surface states from Section~\ref{sec:surface-states}, we can expand
\begin{equation}
	\Psi(\bm{r}) = \frac{1}{\sqrt{\mathcal{A}}}\sum_{\k}e^{i\k\cdot\bm{r}_s}e^{\lambda_{\k} x_b}\Psi_{\k}, 
\end{equation}
where $\bm{r}_s = (x_a,x_c)$ is the position on the surface of the crystal, $\k = (k_a,k_c)$ is a 2D wavevector, $\lambda_{\k}$ describes attenuation into the bulk, and $\Psi_{\k}$ is an electron field operator. Using the polarization field of the SO mode, it has been shown\cite{lucas1970electron, lucas1970quantum, wang1972electron} that the electron-phonon interaction for $q \gg \omega_{TO}/c$ takes the form 
\begin{equation}
	\H_{ep} \propto \sum_{\k,\q}\frac{1}{\sqrt{q}}\left(\int_{-\infty}^0\d x_b\ e^{qx_b}e^{(\lambda_{\k+\q} + \lambda_{\k})x_b} \right)A_{\q}\Psi^{\dag}_{\k + \q}\Psi_{\k}.
\end{equation}
So in contrast with the \citet{frohlich1954} Hamiltonian, where the electron-bulk LO phonon vertex scales like $1/q$, the electron-SO phonon vertex scales like $1/\sqrt{q}$. The coupling of electrons to extended bulk-like LO phonons in a slab geometry was found\cite{lucas1970electron} to scale like the bulk Fr\"ohlich coupling.

\subsection{\label{subsec:density-response} Density Response and Phonon Damping}

\begin{figure}[H]
	\includegraphics[width=\linewidth]{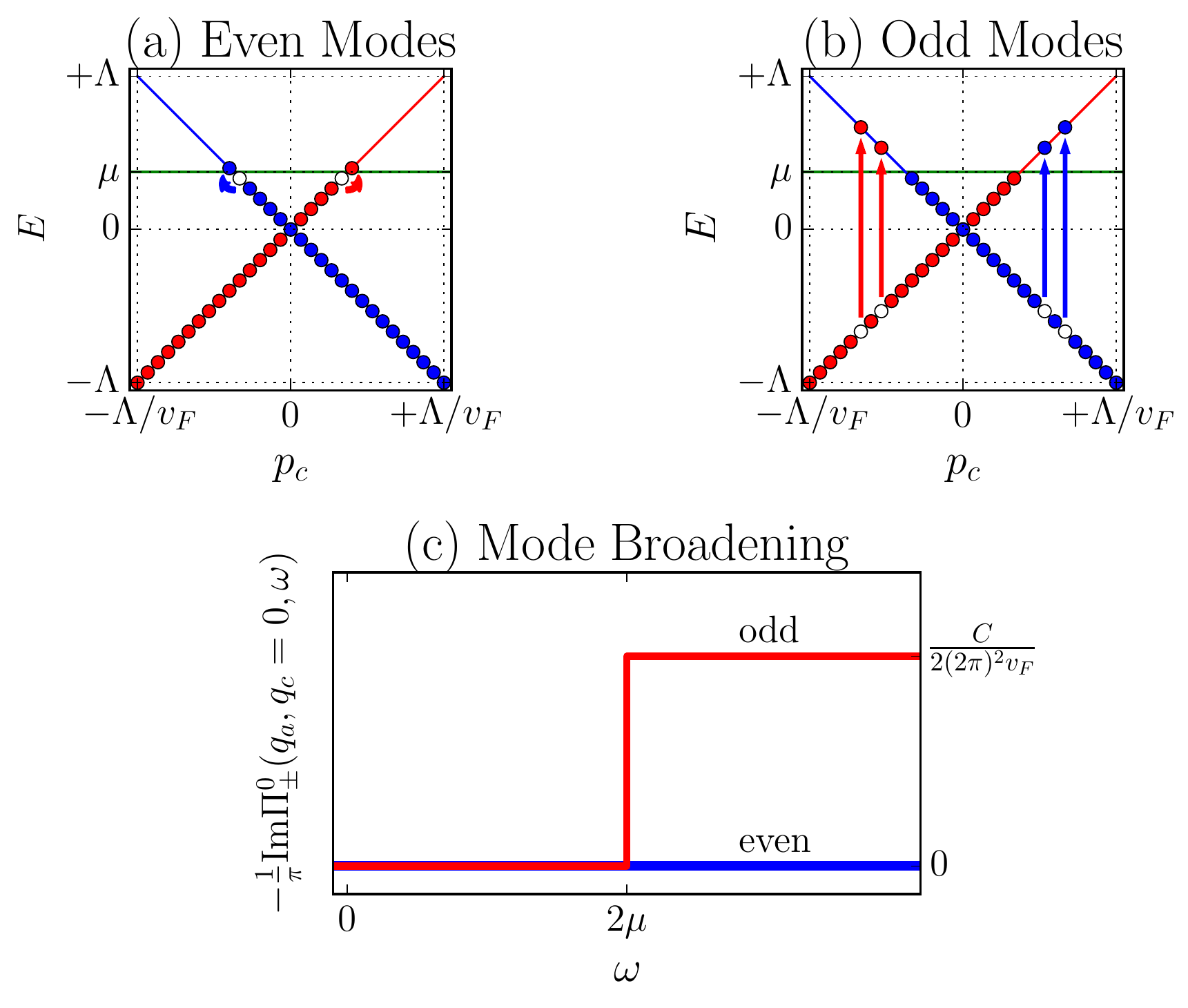}
	\caption{(colour online) Particle-hole excitations contributing to the (a) even mode self-energy for small $q_a$ with $p_c = k_c - \pi$, and (b) the same for odd modes as shown for $k_a > 0$. (c) Imaginary part of phonon self-energy as a function of $\omega$, with $q_c = 0$ and small $q_a$, for even and odd modes.}
	\label{fig:ph-excitations} 
\end{figure}

With the preceding form of the electron-phonon interaction Eq.~\ref{eq:sym-ep-coupling}, we calculate the imaginary part of the first-order polarization bubble $\Pi^0_{\pm}$ in the Matsubara formalism for $q_c = 0$ at zero temperature. The self-energy serves to modify the phonon propagator from Eq.~\ref{eq:unperturbed-propagator} to
\begin{equation}
	\mathcal{D}_{\lambda}(\q,\omega) = \frac{2\omega_{\q\lambda}^0}{\omega^2 - (\omega_{\q\lambda}^0)^2 - 2\omega_{\q\lambda}^0|g_{\q\lambda}|^2\Pi^0_{\lambda}(\q,\omega)}, 
\end{equation}
and in particular, the imaginary part of $\Pi^0_{\lambda}$ broadens the phonon mode lifetime. An explicit calculation (see Appendix~\ref{app:bubble-calculation}) yields 
\begin{equation}
	-\tfrac{1}{\pi}\mathrm{Im}\Pi^0_+(q_a,q_c = 0,\omega) = 0
\end{equation}
for the even modes, and
\begin{equation}
	-\tfrac{1}{\pi}\mathrm{Im}\Pi^0_-(q_a,q_c = 0,\omega) \propto \frac{1}{v_F}\left[\Theta(2\mu + \omega) - \Theta(2\mu - \omega)\right]
\end{equation}
for the odd modes, where $\omega\in [-2\Lambda,+2\Lambda]$ and $q_a$ is small. The chemical potential $\mu > 0$ of the surface states is measured from the nodal point shown in Fig.~\ref{fig:ph-excitations}. The imaginary part is easily understood by considering particle-hole excitations which provide decay channels for the phonons. Even modes excite pairs within the same mirror branch, leading to a linear $\mathrm{Im}\Pi^0_+(\q,\omega)$ as depicted in Fig.~\ref{fig:ph-excitations}(a), which vanishes at $q_c = 0$. Odd modes excite pairs in different mirror branches, so even at $q_c = 0$ pairs can be excited with energy $\omega$ ranging from $2\mu$ to $2\Lambda$ as depicted in Fig.~\ref{fig:ph-excitations}(b).

These results hold for small $q_a$, so long as particle-hole pairs with different $k_a$ lie in the same region along the line of 1D Dirac cones. In real materials the chiral symmetry is slightly broken due to next-nearest neighbour in-layer hopping, which adds a small $k_a$ dispersion. Despite this, particle-hole pairs may still be excited at small $q_a$ with the window $[2\mu,2\Lambda]$ being slightly reduced. We therefore predict damping of mirror odd bulk LO or SO phonons near the zone center due to the presence of surface states, while mirror even phonons are unaffected. 

\section{\label{sec:discussion} Discussion}

Landau damping due to particle-hole excitations of a typical bulk FS, including a nodal ring FS\cite{rhim2016anisotropic}, vanishes as $q\rightarrow 0$ at finite frequency. However, significant damping of particular phonon modes in the same limit occurs in a TCSM, when the TCSM exhibits a set of flat 1D Dirac surface states as shown in Fig.~\ref{fig:SBZ}. Nearly flat bands in one direction is responsible for this effect.

This unique feature of the surface electron-phonon interaction in a TCSM, distinguished from bulk electronic contributions, is associated with the symmetry properties of surface states. With $\q$ along $q_a$, phonons have definite parity under mirror reflection. Even modes can excite electron-hole pairs within the same mirror branch, while odd modes can excite between different branches. As a result, only the odd optical phonons will be damped at the zone center. 

Signatures of this effect may be accessible through a combination of optical and scattering experiments. For SrHfO$_3$, isostructural to SrIrO$_3$, optical mode frequencies have been calculated\cite{vali2009lattice} using density functional perturbation theory (DFPT), which are comparable with experimental Raman\cite{park1976raman,lee2010optical}, and IR reflectivity\cite{lee2010optical} studies. In the case of SrIrO$_3$, high pressures are necessary to achieve the orthorhombic perovskite structure\cite{longo1971structure}, and as a consequence there are only a small number of bulk experiments available\cite{nie2015interplay, matsuno2015engineering, zhao2008high,blanchard2014anomalous,fujioka2017correlated}. Despite this, line widths in bulk Raman spectroscopy, or the imaginary part of the dielectric function from IR reflectivity, should reflect electronic damping. Since SrHfO$_3$ is is electronically insulating it may serve as a reference material for intrinsic line widths. 

Moreover, Fuchs-Kliewer SO modes have been observed\cite{baden1981observation} in the cubic SrTiO$_3$ through high-resolution, low-energy electron diffraction (LEED). Based on the general theory of these modes, we also expect them to exist near the zone center in SrIrO$_3$ between bulk LO and TO frequencies, provided that the dielectric function is negative. Bulk IR reflectivity data would serve to determine $\epsilon(\omega)$ as well as the LO and TO frequencies. Even without a microscopic basis for the SO modes (which may be provided by DFPT, or a semi-empirical approach such as the embedded atom method\cite{daw1984embedded,daw1993embedded,foiles1986embedded,karimi1992embedded} or the multipole expansion\cite{jayanthi1987nature,kaden1992electronic}), a general symmetry analysis tells us that SO modes along $q_a$ will have definite parity under mirror reflection. The existence of these modes could be confirmed with LEED or inelastic helium atom scattering (HAS)\cite{zhu2011interaction,zhu2012epcoupling}. In SrIrO$_3$ HAS would be more appropriate to avoid electronic contributions to scattering. Comparing line widths in time-of-flight HAS measurements of SrIrO$_3$ along the $\hat{\bm{a}}$ direction (for a $\hat{\bm{b}}$ crystal termination) with the reference material SrHfO$_3$ would provide information about how the SO modes are damped. Quantitative analysis of SO modes in SrIrO$_3$ is beyond the scope of the current work, and may be an interesting subject for future study. 

In summary, we have investigated the nature of the wave function describing surface states in a TCSM, and found unique properties under mirror symmetry. This restricts the form of the electron-phonon interaction for phonons of definite mirror symmetry when $\q$ is along $q_a$. The surface states couple to bulk LO, or SO modes with different scaling of the vertex. We computed the first-order self-energy of mirror even and odd phonons, and found that damping near the zone center at finite frequency is zero for mirror even modes but finite for mirror odd modes. Damping from surface electrons is distinct from typical Landau damping due to bulk electrons, and we propose a combination of optical and HAS experiments to observe this. Such an experiment would provide the first evidence of surface states in a topological crystalline semimetal.

\begin{acknowledgments}
We thank Yige Chen for useful discussions at the beginning of this project. This work was supported by the Natural Sciences and Engineering Research Council of Canada and the Center for Quantum Materials at the University of Toronto.
\end{acknowledgments}

\appendix
\renewcommand\theequation{A\arabic{equation}}
\section{\label{app:tb-hamiltonian} Tight-Binding Hamiltonian of SrIrO$_3$}

The $j_{\mathrm{eff}} = \tfrac{1}{2}$ tight-binding model Eq.~\ref{eq:full-tb} of SrIrO$_3$ is defined in terms of the following $\k$ dependent functions
\begin{align}\label{eq:tb-functions}
\begin{split}
\lambda_{\k} &= \tfrac{1}{2}t_{xy}[\cos(k_a) + \cos(k_b)] \\
\E^p_{\k} &= 2(2t_p - it_p')\cos(\tfrac{1}{2}k_a)\cos(\tfrac{1}{2}k_b) \\
\E^z_{\k} &= 2t_z\cos(\tfrac{1}{2}k_c) \\
\E^d_{\k} &= 2t_d\cos(\tfrac{1}{2}k_a)\cos(\tfrac{1}{2}k_b)\cos(\tfrac{1}{2}k_c) + \\
&\hspace{5mm} 2it_d'\sin(\tfrac{1}{2}k_a)\cos(\tfrac{1}{2}k_b)\sin(\tfrac{1}{2}k_c)\\
\E^{po}_{\k} &= (1 - i)(t_{1p}^o + t_{2p}^o)\cos(\tfrac{1}{2}k_a)\cos(\tfrac{1}{2}k_b) + \\
&\hspace{5mm} (1 + i)(t_{1p}^o - t_{2p}^o)\sin(\tfrac{1}{2}k_a)\sin(\tfrac{1}{2}k_b) \\
\E^{zo}_{\k} &= t_z^o(1 - i)\cos(\tfrac{1}{2}k_c) \\
\E^{do}_{\k} &= (1 + i)t_d^o\sin(\tfrac{1}{2}k_a)\cos(\tfrac{1}{2}k_b)\sin(\tfrac{1}{2}k_c) + \\
&\hspace{5mm}   (1 - i)t_d^o\cos(\tfrac{1}{2}k_a)\sin(\tfrac{1}{2}k_b)\sin(\tfrac{1}{2}k_c) \\
\E^{d1}_{\k} &= (1 + i)t_d^1\cos(\tfrac{1}{2}k_a)\cos(\tfrac{1}{2}k_b)\cos(\tfrac{1}{2}k_c) - \\
&\hspace{5mm} (1 - i)t_d^1\sin(\tfrac{1}{2}k_a)\sin(\tfrac{1}{2}k_b)\cos(\tfrac{1}{2}k_c),
\end{split}
\end{align}
where $\k$ is written in the orthorhombic basis $(k_a,k_b,k_c)$, and $\{t_{xy},t_p,t_p',t_z,t_d,t_d',t_{1p}^o,t_{2p}^o,t_z^o,t_d^o,t_d^1\}$ are tight-binding parameters with values listed in Table~\ref{table:tb-parameters}. Three sets of Pauli matrices $\bm{\sigma},\bm{\nu},\bm{\tau}$ are used to define the Hamiltonian, and have the following effect on the pseudospins $\uparrow,\downarrow$ and the iridium sites $B,R,Y,G$
\begin{equation}
	\uparrow\ \stackrel{\sigma_x}{\longleftrightarrow}\ \downarrow \qquad 
	\begin{array}{cc}
		B\ \stackrel{\nu_x}{\longleftrightarrow}\ Y &\qquad R\ \stackrel{\nu_x}{\longleftrightarrow}\ G \\
		B\ \stackrel{\tau_x}{\longleftrightarrow}\ R &\qquad Y\ \stackrel{\tau_x}{\longleftrightarrow}\ G.
	\end{array}
\end{equation}

\begin{table}
\caption{\label{table:tb-parameters} Tight-binding parameters for SrIrO$_3$ in eV.\cite{carter2012semimetal,chen2015topological}}
\begin{ruledtabular}
\begin{tabular}{c  c c c c c c c c c c}
$t_{xy}$ & $t_p$ & $t_p'$ & $t_z$ & $t_d$ & $t_d'$ & $t_{1p}^o$ & $t_{2p}^o$ & $t_z^o$ & $t_d^o$ & $t_d^1$ \\
\hline 
$-0.3$ & $-0.6$ & $-0.15$ & $-0.6$ & $-0.3$ & $0.03$ & $0.1$ & $0.3$ & $0.13$ & $0.06$ & $0.03$
\end{tabular}
\end{ruledtabular}
\end{table}

\renewcommand\theequation{B\arabic{equation}}
\section{\label{app:bubble-calculation} Calculation of \texorpdfstring{$-\tfrac{1}{\pi}\mathrm{Im}\Pi^0_{\pm}(\bm{q},\omega)$}{$-\tfrac{1}{\pi}\mathrm{Im}\Pi^0_{\pm}(\vec{q},\omega)$}}

In this section we calculate the imaginary part of the first order polarization bubbles $\Pi^0_{\pm}(\bm{q},\omega)$ using the Matsubara formalism, shown diagrammatically in Fig.~\ref{fig:phonon-RPA} with $q = (\q,iq_m)$. The form of the bubbles are\cite{mahan}
\begin{equation}
	\Pi^0_{\pm}(q) = \frac{1}{\beta\A}\sum_{\k,ik_n}\mathrm{Tr}[\G^0(k)\Gamma_{\pm}\G^0(k + q)\Gamma_{\pm}],
\end{equation}
where the trace is taken over branches of the surface states, and $\Gamma_{\pm}$ enforces total and relative density respectively ($\id$, and $\gamma_x$ in the basis of the surface states). The electron Matsubara Green's function is
\begin{equation}
	\G^0 = \begin{pmatrix}\G^0_+ &  \\  & \G^0_-\end{pmatrix} \quad \G^0_m(\k,ik_n) = \frac{1}{ik_n - \xi_{\k m}}, 
\end{equation}
where $\xi_{\k m} = \varepsilon_{\k m} - \mu$, and $\mu$ is the chemical potential measured from the nodal point. As the dispersion we take Eq.~\ref{eq:ss-dispersion}
$$\varepsilon_{\k m} = m\cdot v_F\cdot  \mathrm{sgn}(-k_a)\cdot(k_c - \pi) \qquad |k_c - \pi| \le \Lambda/v_F,$$
for some high-energy cutoff $\Lambda$.  

Starting with $\Pi^0_+$, after the Matsubara sum, performing the analytic continuation, and passing to the continuum limit, at $T = 0$ we obtain
\begin{widetext}
\begin{equation}
	\Pi^0_+(\q,\w) = \sum_{m}\int\frac{\d\k}{(2\pi)^2}\frac{\Theta(-\xi_{\k m}) - \Theta(-\xi_{\k + \q m})}{\w + \xi_{\k m} - \xi_{\k+\q m} + i\eta} 
				   = \sum_{m}\int\frac{\d\k}{(2\pi)^2}\frac{\Theta(-\xi_{\k m}) - \Theta(-\xi_{\k + \q m})}{\w + mv_F\mathrm{sgn}(k_a)q_c + i\eta},
\end{equation}
\end{widetext}
assuming $q_a$ is sufficiently small such that particle-hole pairs remain within one $k_a$ region. As a consequence of $q_a$ being small the slight $k_a$ dispersion introduced by breaking of the chiral symmetry will have little effect on our results. We set $q_c = 0$ so that the phonon mode does not break the mirror symmetry, which causes the Fermi factors to cancel as the dispersion is flat in the $k_a$ direction.  Therefore, the imaginary part of $\Pi^0_+$ vanishes
\begin{equation}
	-\tfrac{1}{\pi}\mathrm{Im}\Pi^0_+(q_a, q_c = 0,\w) = 0,
\end{equation}
for small $q_a$. 

Next, we focus on $\Pi^0_-$ which contributes
\begin{align}
\begin{split}\label{eq:pio-m}
	\Pi^0_-(\q,\w) &= \sum_{m}\int\frac{\d\k}{(2\pi)^2}\frac{\Theta(-\xi_{\k m}) - \Theta(-\xi_{\k + \q \overline{m}})}{\w + \xi_{\k m} - \xi_{\k+\q \overline{m}} + i\eta} \\
				   &= \frac{C}{(2\pi)^2}\sum_{m}\int_{-\Lambda/v_F}^{+\Lambda/v_F}\d p_c\frac{\Theta(-\xi_{\k m}) - \Theta(-\xi_{\k + \q \overline{m}})}{\w - 2mv_Fp_c + i\eta},
\end{split}
\end{align}
where we have again assumed small $q_a$, focused on $k_a > 0$ ($k_a < 0$ contributes identically), set $q_c = 0$, and shifted the $k_c$ integration to $p_c = k_c - \pi$. The $k_a$ integral has been performed which contributes an overall constant $C$, which is approximately the width of the $k_a$ region of validity set by the cutoff $\Lambda$. With the estimated region from Section~\ref{subsec:slab-calculations}, $C \approx 0.7\pi $. Taking the imaginary part of this quantity, the delta function fixes
\begin{equation}
	2mv_Fp_c = \omega,
\end{equation}
and constrains $\omega$ to lie in the range $[-2\Lambda,+2\Lambda]$.
The Fermi factors are
\begin{equation}
\Theta(2\mu  + \omega) - \Theta(2\mu - \omega).
\end{equation}
\hspace{-3pt}For each $m$, the above factors combine to form
\begin{equation}
	\frac{C}{2v_F(2\pi)^2}\mathrm{sgn}(\omega)\begin{cases}1 & 2\mu \le |\omega| \le  2\Lambda \\ 0 & \mathrm{else}\end{cases}.
\end{equation}
Combining these contributions, we find
\begin{align}
\begin{split}
	-\tfrac{1}{\pi}\mathrm{Im}\Pi^0_-(q_a,\omega) &= \frac{C}{v_F(2\pi)^2}\left[\Theta(2\mu + \omega) - \Theta(2\mu - \omega)\right]\\ 
												  &= \frac{C}{v_F(2\pi)^2}\mathrm{sgn}(\omega) \begin{cases}1 & 2\mu \le |\omega| \le  2\Lambda \\ 0 & 																			\mathrm{else}\end{cases}
\end{split}
\end{align}
for $q_c = 0$ and small $q_a$. A sketch of the relevant particle-hole excitations contributing to $-\tfrac{1}{\pi}\mathrm{Im}\Pi^0_{\pm}$ are shown in Fig.~\ref{fig:ph-excitations}.

The real part of $\Pi_-^0$ can also be calculated directly from Eq.~\ref{eq:pio-m}. Focusing on $k_a > 0$, the Fermi factors are $\Theta(\mu + mv_Fp_c) - \Theta(\mu - mv_Fp_c)$ which simplify to $m \mathrm{sgn}(p_c)\Theta(|p_c| - \mu/v_F)$. The terms with $m = \pm$ contribute identically, so we will focus on $m = +$. For $p_c > 0$ we have the term
\begin{align}
\begin{split}
	I_1 &=  \frac{C}{(2\pi)^2}\mathcal{P}\int_{\mu/v_F}^{\Lambda/v_F}\frac{\d p_c}{\omega - 2v_Fp_c} \\
		&= \frac{C}{2v_F(2\pi)^2}\mathcal{P}\int_{\Lambda/v_F}^{\mu/v_F}\frac{\d p_c}{p_c - \omega/2v_F} \\
		&= \frac{C}{2v_F(2\pi)^2}\log\left|\frac{\omega - 2\mu}{\omega - 2\Lambda}\right|,
\end{split}
\end{align}
where $\mathcal{P}$ denotes the Cauchy principal value. Similarly, for $p_c < 0$ we have the term 
\begin{align}
\begin{split}
	I_2 &=  \frac{C}{(2\pi)^2}\mathcal{P}\int_{-\Lambda/v_F}^{-\mu/v_F}\frac{\d p_c}{2v_Fp_c - \omega} \\
		&= \frac{C}{2v_F(2\pi)^2}\mathcal{P}\int_{-\Lambda/v_F}^{-\mu/v_F}\frac{\d p_c}{p_c - \omega/2v_F} \\
		&= \frac{C}{2v_F(2\pi)^2}\log\left|\frac{\omega + 2\mu}{\omega + 2\Lambda}\right|.
\end{split}
\end{align}
Therefore, $\mathrm{Re}\Pi_-^0 = 2(I_1 + I_2)$ evaluates to
\begin{equation}
	\mathrm{Re}\Pi_-^0(q_a,\omega) = \frac{C}{v_F(2\pi)^2}\log\left|\frac{(\omega - 2\mu)(\omega + 2\mu)}{(\omega - 2\Lambda)(\omega + 2\Lambda)}\right|.
\end{equation} \\[1mm]
One can obtain the same result using the Kramers-Kronig relation\cite{mahan}
\begin{equation}
	\mathrm{Re} \Pi_-^0(q_a,\omega) = \mathcal{P}\int_{-\infty}^{+\infty}\frac{\d\omega'}{\omega - \omega'}\left[-\frac{1}{\pi}\mathrm{Im}\Pi_-^0(q_a,\omega')\right].
\end{equation}
\vfill 

\bibliography{references}

\end{document}